\documentclass[twocolumn,pra,aps,superscriptaddress]{revtex4-2}
\usepackage{amsmath}
\usepackage{amssymb}
\usepackage{amsfonts}
\usepackage[dvips]{graphicx}
\usepackage{subfigure}
\usepackage{dcolumn}
\usepackage{txfonts}
\usepackage{bm}
\usepackage{makeidx}
\usepackage{color}
\usepackage{mathtools}
\usepackage{threeparttable}
\usepackage[colorlinks,linkcolor=blue,anchorcolor=blue,citecolor=blue,urlcolor=blue]{hyperref}

\begin{document}

\title{Duality Symmetry of Quantum Electrodynamics and Magnetic Charges}
\author{Li-Ping Yang}

% \email{lipingyang87@gmail.com}
\affiliation{Center for Quantum Sciences and School of Physics, Northeast Normal University, Changchun 130024, China}

\author{Dazhi Xu}
\affiliation{Center for Quantum Technology Research and Key Laboratory of Advanced Optoelectronic Quantum Architecture and Measurement (MOE) and School of Physics, Beijing Institute of Technology, Beijing 100081, China}

\begin{abstract}
The duality symmetry between electricity and magnetism hidden in classical Maxwell equations suggests the existence of dual charges, which have usually been interpreted as magnetic charges and have not been observed in experiments. In quantum electrodynamics (QED), both the electric and magnetic fields have been unified into one gauge field, which makes this symmetry inconspicuous. Here, we recheck the duality symmetry of QED by introducing a dual gauge field and a dual symmetric Lagrangian. Within the framework of gauge-field theory, we show that the electric-magnetic duality symmetry cannot give any new conservation law. By checking the charge-charge interaction and the quantum Lorentz-force equation, we find that the introduced dual charges are electric charges, not magnetic charges. More importantly, we show that true magnetic charges are not compatible with the gauge-field theory of QED, because the interaction between a magnetic charge and an electric charge cannot be mediated via the exchange of gauge photons.
\end{abstract}

\maketitle
\section{Introduction}
\begin{figure*}
\centering
\includegraphics[width=15cm]{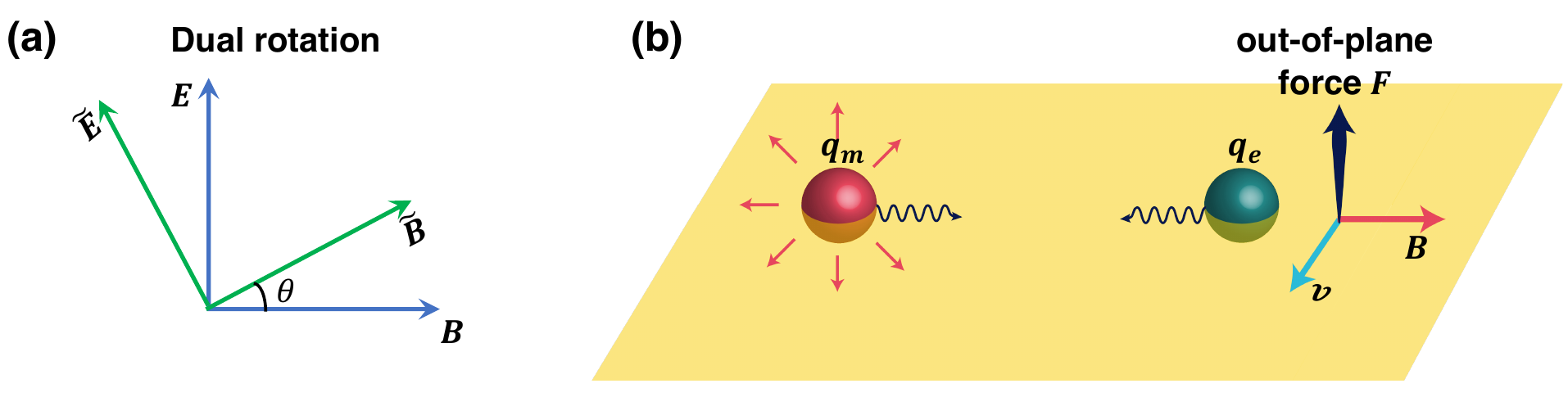}\caption{\label{fig:schematic} (a) Dual rotation between electric and magnetic fields. (b) Interaction between a magnetic charge and an electric charge  mediated by exchanging gauge photons. In the Coulomb-like magnetic field generated by a static magnetic charge, a moving electric charge will experience an out-of-plane force, which will violate the conservation of momentum in the microscopic photon exchanging processes.}
\end{figure*}

Maxwell equations for classical electrodynamics (CED) exhibit the high symmetry between electricity and magnetism after a dual rotation between electric and magnetic fields~\cite{jackson1999classical} [see Fig.~\ref{fig:schematic} (a)]. This electric-magnetic duality symmetry suggests the existence of dual charges, which have been regarded as magnetic charges usually~\cite{Strazhev1972dual}. However, the physical nature of the dual charge in symmetric Maxwell equations [see Eq.~(\ref{eq:Maxwell})] has not been fully clarified. The microscopic details of the magnetic-charge-related electromagnetic interaction remain unclear.  Specifically, the role of gauge photons in mediating the interaction between a magnetic charge and an electric charge is unknown.

A true magnetic charge is defined as the source of a static Coulomb-like magnetic field. The Lorentz-force equation shows that the trajectory of an electron moving in a Coulomb-like magnetic field
lies on an axially symmetric Poincar\'e cone~\cite{shnir2006magnetic}. Thomson found that a stationary dipole formed by an electric charge $q_e$ and a magnetic charge $q_m$ carries non-vanishing angular momentum due to a non-trivial contribution from the Coulomb-like fields~\cite{thomson1904momentum,jackson1999classical,shnir2006magnetic}.  Some of the experiments in searching magnetic charges are also based on classical Maxwell equations~\cite{Patrizii2015status,mavromatos2020magnetic}, such as magnetic-flux-measurement-based-methods~\cite{Cabrera1982first,Bai2021searching},  \v{C}erenkov-emission based method~\cite{Tompkins1965total,Abbasi2022search}. However, we show that the gauge-field theory gives a significantly different picture to understand the basic interaction between magnetic charges and electromagnetic (EM) fields.

The duality symmetry of source-free transverse EM fields has been exploited to explain the conservation of photon helicity recently~\cite{cameron2012electric,bliokh2013dual,drummond1999dual,Elbistan2017Duality}. This conservation law reveals the physical nature of the conserved quantity discovered by Lipkin previously~\cite{Lipkin1964existence,calkin1965invariance,candlin1965analysis}. The corresponding continuity equation also creates the link between the photon helicity density and the photon spin density~\cite{barnett2012duplex}. Deser and Teitelboirn showed that duality rotations of electric and magnetic fields for source-free Maxwell theory can be implemented by a time-local generator  in an arbitrary spacetime metric~\cite{Deser1976duality}.  However, in the absence of charges, electric and magnetic fields are physically indistinguishable. Charges are essential to understanding the duality symmetry. The light-charge interaction also plays a significant role in obtaining gauge-invariant angular momentum observables of QED~\cite{yang2020quantum}. On the other hand, both electric and magnetic fields have been unified within one gauge field in the standard QED theory. The electric-magnetic duality symmetry of QED remains elusive.

We now recheck the duality symmetry of QED by introducing a symmetric Lagrangian with an extra gauge field and dual charges. We find no new conservation law can be obtained from the electric-magnetic duality symmetry. By investigating the microscopic charge-charge interaction, we show that quantum gauge-field theory exhibits a significant departure from the classical Maxwell's theory. According to the classical Lorentz-force equation, a moving electric charge in the Coulomb-like magnetic field of a magnetic charge [see Fig.~\ref{fig:schematic} (b)] will experience an out-of-plane force. However, this force cannot be mediated by exchanging gauge photons. We derive the quantum Lorentz-force equation to elucidate this discrepancy. We conclude that the introduced dual charges from the duality symmetry are actually electric charges, not magnetic charges~\cite{Strazhev1972dual}. Magnetic charges are compatible with classical Maxwell equations, but not the QED gauge-field theory, which is the most accurate and successful theory in physics today. A similar conclusion has been reached by Weinberg that \textit{it is impossible to construct a Lorentz-invariant $S$ matrix for magnetic monopoles and charges in perturbation theory}~\cite{Weinberg1965photons} and also by Zwanziger~\cite{Zwanziger1965Dirac}.

We note that the magnetic charge involved in this work is not the Dirac monopole, which is composed of a magnetic charge and a nodal line with diverging gauge potential and quantized flux~\cite{Dirac1930Quantised,Dirac1948monopole}. Dirac remarkably suggested that the mere existence of one Dirac monopole will lead to the quantization of all electric charges. Thus, even though has not been observed in experiments, this hypothesized particle has attracted intensive interests~\cite{Schwinger1966magnetic,Zwanziger1968quantum,Hooft1974magntic,polyakov1974particle,wu1975concept,singleton1996electromagnetism,goddard1978magnetic,weinberg2007magnetic}. In the following, we limit ourselves only to the topologically trivial magnetic charges for EM fields. Our results may offer a different perspective on magnetic monopoles problems. %Wu and Yang showed the Dirac monopole corresponds to a topologically non-trivial solution of U(1) gauge fields and generalized Dirac's monopole concept for non-Abelian gauge fields~\cite{wu1975concept}.  ’t Hooft and Polyakov discovered regular magnetic monopoles for non-Abelian gauge fields~\cite{Hooft1974magntic,polyakov1974particle}.

\section{Duality symmetry of classical electrodynamics}
In this section, we give a brief review of the duality symmetry of the CED. After a dual transformation shown in Fig.~\ref{fig:schematic} (a), we can rewrite the Maxwell equations in a highly symmetric form~\cite{jackson1999classical}
\begin{equation}
\begin{split}
\boldsymbol{\nabla}\cdot\tilde{\boldsymbol{E}}& =\frac{\tilde{\rho}_e}{\varepsilon_0},\ \boldsymbol{\nabla}\times \tilde{\boldsymbol{B}}=\frac{1}{c^2}\frac{\partial }{\partial t}\tilde{\boldsymbol{E}} +\mu_0\tilde{\boldsymbol{J}}_e, \\
\boldsymbol{\nabla}\cdot\tilde{\boldsymbol{B}}& =\tilde{\rho}_m,\ -\boldsymbol{\nabla}\times \tilde{\boldsymbol{E}}= \frac{\partial }{\partial t}\tilde{\boldsymbol{B}} +\tilde{\boldsymbol{J}}_m, 
\end{split}\label{eq:Maxwell}      
\end{equation}
where the vacuum light speed $c=1/\sqrt{\varepsilon_0\mu_0}$ is determined by the vacuum permittivity $\varepsilon_0$ and permeability $\mu_0$. The new quantities $\tilde{\rho}_m$ and $\tilde{\boldsymbol{J}}_m$ are the charge density and current corresponding to the dual charge $\tilde{q}_m$, which has been interpreted as the magnetic charge usually. We will re-check this magnetic-charge interpretation within the QED gauge-field framework in the following. Without causing confusion, we will call  $\tilde{q}_m$ from the dual transformation as the dual charge.  The superscript $\tilde{\ }$ has been added to distinguish this representation from the asymmetric one in the absence of dual charges. 

In addition to the traditional four-vector potential $\tilde{A}^{\mu}=(\tilde{A}^0,\tilde{\boldsymbol{A}})$, we can also introduce a dual four-vector potential $\tilde{C}^{\mu}=(\tilde{C}^0,\tilde{\boldsymbol{C}})$ to re-express the electric and magnetic fields into a symmetric form~\cite{singleton1996electromagnetism} 
\begin{align}
\tilde{\boldsymbol{E}} & =-(c\partial^{0}\tilde{\boldsymbol{A}}+c\boldsymbol{\nabla}\tilde{A}^{0}+\boldsymbol{\nabla}\times\tilde{\boldsymbol{C}}),\label{eq:E}\\
\tilde{\boldsymbol{B}} & =-(\partial^{0}\tilde{\boldsymbol{C}}+\boldsymbol{\nabla}\tilde{C}^{0}-c\boldsymbol{\nabla}\times\tilde{\boldsymbol{A}})/c.\label{eq:B}
\end{align}
We will also have two EM tensors $\tilde{F}^{\mu\nu}=\partial^{\mu}\tilde{A}^{\nu}-\partial^{\nu}\tilde{A}^{\mu}$ and $\tilde{G}^{\mu\nu}=\partial^{\mu}\tilde{C}^{\nu}-\partial^{\nu}\tilde{C}^{\mu}$ in the symmetric representation. We note that the new EM tensor $\tilde{G}^{\mu\nu}$ is not the conventional dual tensor $\mathcal{F}^{\mu\nu}=\epsilon^{\mu\nu\alpha\beta}F_{\alpha\beta}/2$ ($\epsilon^{\mu\nu\alpha\beta}$ is the rank four Levi-Civita tensor)~\cite{singleton1996electromagnetism}, which can be obtained by the special dual transformation with $\theta=3\pi/2$ as shown in Fig.~\ref{fig:schematic} (a), i.e., $\boldsymbol{E}\rightarrow c\boldsymbol{B}$ and $\boldsymbol{B}\rightarrow -\boldsymbol{E}/c$. We can return to the conventional asymmetric representation via inverse dual transformations
\begin{align}
\boldsymbol{E} & = \tilde{\boldsymbol{E}}\cos\theta +c\tilde{\boldsymbol{B}}\sin\theta,\ \boldsymbol{B} = \tilde{\boldsymbol{B}}\cos\theta -(\tilde{\boldsymbol{E}}/c)\sin\theta,\label{eq:Transform1}\\   
q_e & \!=\! \tilde{q}_e\! \cos\theta \!+\!c\varepsilon_0\tilde{q}_m\sin\theta,\ q_m \!= \!\tilde{q}_m\! \cos\theta\!-\! (\tilde{q}_e/c\varepsilon_0)\!\sin\theta,\label{eq:Transform2}\\
A^{\mu} &\! =\! \tilde{A}^{\mu}\cos\theta \!+\!(\tilde{C}^{\mu}/c)\sin\theta,\ C^\mu\! =\! \tilde{C}^{\mu}\cos\theta \!-\!c\tilde{A}^{\mu}\sin\theta.\label{eq:Transform3}
\end{align}
The transformation of $\rho_{e(m)}$ or the components of $\boldsymbol{J}_{e(m)}$ is the same as $q_{e(m)}$. We note that the dual rotation angle $\theta$ is determined by the ratio of the two charges, i.e., $\tan\theta=c\varepsilon_0\tilde{q}_m/\tilde{q}_e$, such that after the inverse transformation, all the dual charge related quantities (i.e., $q_m$, $\rho_m$, $\boldsymbol{J}_m$, etc) vanish in the asymmetric representation.

To guarantee the equivalence of these two representations~\cite{jackson1999classical}, we need to require that all charge particles have the same ratio between dual charge $\tilde{q}_m$ and electric charge $\tilde{q}_e$. The total charge (the norm) $q=\sqrt{c^2\varepsilon_0^2\tilde{q}^2_m+\tilde{q}^{2}_e}$ keeps invariant under a dual rotation. On the other hand, if all charged particles have the same magnetic/electric charge ratio, we can always transform to an asymmetric representation with only one type of charge~\cite{jackson1999classical,Strazhev1972dual}. Thus, strictly speaking, not magnetic charges but two charges with different magnetic/electric ratios have not been observed in experiments. We emphasize that the dual charge $\tilde{q}_m$ is significantly different from Dirac's magnetic monopole, which was utilized to explain the fundamental origin of charge quantization~\cite{Dirac1948monopole}. Here, $\tilde{q}_m$ can be an arbitrary portion of the quantized total charge $q$. The dual charge cannot give the quantization condition of the charges. 

On the other hand, a subsidiary condition for the two gauge fields $\tilde{q}_e\tilde{C}^{\mu}-\tilde{q}_m\tilde{A}^{\mu}/\mu_0=0$~\cite{shnir2006magnetic}, i.e.,
\begin{equation}
\tilde{C}^{\mu}\cos\theta=c\tilde{A}^{\mu}\sin\theta, \label{eq:Subsidary}   
\end{equation}
has to be enforced to obtain vanishing $C^{\mu}=0$ and EM tensor $G^{\mu\nu}=0$ in the asymmetric representation. The gauge field $\tilde{C}^{\mu}$ is significantly different from the one having been introduced to construct the dual EM field tensor previously~\cite{cameron2012electric,bliokh2013dual}. The subsidiary condition clearly shows that the two gauge fields $\tilde{A}^{\mu}$ and $\tilde{C}^{\mu}$ are not independent variables and they have to share the same gauge freedom.  

In additional to Maxwell equations, the empirical Lorentz-force equation $\boldsymbol{F}=q_e(\boldsymbol{E}+\boldsymbol{v}\times\boldsymbol{B})$ is required for the complete description of CED. After a dual transformation, we obtain its counterpart in the symmetric representation,
\begin{equation}
\boldsymbol{F}=\frac{d}{dt}\boldsymbol{p}_{\rm mech}=\tilde{q}_{e}\left(\tilde{\boldsymbol{E}}+\boldsymbol{v}\times\tilde{\boldsymbol{B}}\right)+c\varepsilon_{0}\tilde{q}_{m}\left(c\tilde{\boldsymbol{B}}-\frac{1}{c}\boldsymbol{v}\times\tilde{\boldsymbol{E}}\right).\label{eq:Force_C}
\end{equation}
Here, $\boldsymbol{p}_{\rm mech}=m\boldsymbol{v}$ is the mechanical momentum of a particle with velocity $\boldsymbol{v}$. The time component of the Lorentz force equation is not related to our discussion in this work, so we will not analyze it further. 

Up to now, all results have been obtained from the dual rotation of their counterparts in the original asymmetric representation. Thus, these two representations should exactly equivalent to each other. However, we note that Maxwell wrote his equations in an asymmetric form only because no magnetic charge has been observed in experiments. Magnetic charges, which are defined as the sources of static Coulomb-like magnetic fields, are completely compatible with Maxwell equations. In a world with electric-magnetic duality symmetry, specifically with both magnetic and electric charges, the Maxwell equations, EM fields, EM tensor, and Lorentz force equation should be in their symmetric forms presented here, because they are invariant under a dual rotation. In this case, a moving electric charge will experience an out-of-plane force in the magnetic field generated by a static magnetic charge as shown in Fig.~\ref{fig:schematic} (b).

There are three important problems having not been clarified in CED. Firstly, what is the fundamentally new conservation law corresponding to the duality symmetry? Secondly, will the quantized gauge field $\tilde{C}^{\mu}$ lead to new gauge bosons (i.e., "magnetic" photons~\cite{singleton1996electromagnetism}) that has not been observed? Thirdly, the dual charge $\tilde{q}_m$ has been confused with the Dirac monopole historically and its fundamental nature has not been revealed. Specifically, the microscopic mechanism of the interaction between a dual charge and an electric charge is still missing. Next, we will answer these questions conclusively under the gauge-field framework of QED.  

\section{Duality symmetry of quantum electrodynamics}
The previous section shows that CED based on Maxwell equations can restore its duality symmetry via a dual rotation~\cite{jackson1999classical}. However, modern quantum field theory shows that U(1) gauge-field theory provides a more fundamental framework for electrodynamics, specifically for QED. We note that Maxwell equations and the U(1) gauge-field theory are not exactly equivalent to each other, because gauge fields include extra gauge-dependent degrees of freedom, such as the not directly observable scalar photons and longitudinally polarized photons~\cite{greiner2013field,cohen1997photons}. The Aharonov-Bohm effect~\cite{Aharonov1959significance}, which is a purely quantum effect, cannot be explained with the classical theory based on the local interaction between charges and EM fields. Recently, the scalar and longitudinally polarized photons have also shown to be essential to construct the full spin and orbital angular momentum operator of light~\cite{yang2020quantum}. Thus, the duality symmetry rooting in classical Maxwell equations has to be properly checked under the U(1) gauge-field framework. 

We start from the QED Lagrangian density in the symmetric representation $\tilde{\mathcal{L}}_{\rm QED}=\tilde{\mathcal{L}}_{D}+\tilde{\mathcal{L}}_{M}+\tilde{\mathcal{L}}_{{\rm int}}$, with unchanged Lagrangian density for the Dirac field $\tilde{\mathcal{L}}_{D}=\mathcal{L}_{D}=i\hbar c\bar{\psi}\gamma^{\mu}\partial_{\mu}\psi-mc^{2}\bar{\psi}\psi$. The Fermi Lagrangian density for the EM field changes to
\begin{equation}
 \tilde{\mathcal{L}}_{M} =-\frac{1}{2}\left[ \frac{1}{\mu_{0}}(\partial_{\mu}\tilde{A}^{\nu})(\partial^{\mu}\tilde{A}_{\nu})+\varepsilon_{0}(\partial_{\mu}\tilde{C}^{\nu})(\partial^{\mu}\tilde{C}_{\nu})\right],  
\end{equation}
and the interaction part is given by
\begin{equation}
\mathcal{\tilde{L}}_{{\rm int}}=-c\bar{\psi}\gamma_{\mu}(\tilde{q}_{e}\tilde{A}^{\mu}+\varepsilon_{0}\tilde{q}_{m}\tilde{C}^{\mu})\psi. \label{eq:L_int}  
\end{equation}
Here, the subscript $_{M}$ denotes Maxwell. As shown in Appendix~\ref{sec:DualLagrangian}, our Lagrangian density $\tilde{\mathcal{L}}_{M}$ is obtained via a dual rotation of its conventional counterpart in the asymmetric representation directly. This is significantly different from previous literature~\cite{cameron2012electric,bliokh2013dual}, in which the Lagrangian itself vanishes. The connection between our dual-symmetric Lagrangian and  the symmetric Maxwell equations is shown in Appendix~\ref{sec:MaxwellEqs}. We note that the subsidiary condition has been used in deriving the symmetric Lagrangian. However, the symmetric Lagrangian $\tilde{\mathcal{L}}_{\rm QED}$ itself is invariant under a dual transformation even without the subsidiary condition. This subsidiary condition is only needed in building the connection to the original asymmetric representation.

We now apply Noether's theorem on the duality symmetry to derive the corresponding conservation law~\cite{greiner2013field}. The infinitesimal dual
symmetry transformation can be written as
\begin{align}
x_{\mu}^{\prime} & =x_{\mu},\ 
\psi' =\psi,\\
\tilde{A}_{\mu}^{\prime} & =\tilde{A}_{\mu}+\delta\tilde{A}_{\mu}=\tilde{A}_{\mu}+\theta\tilde{C}_{\mu}/c\\
\tilde{C}_{\mu}^{\prime} & =\tilde{C}_{\mu}+\delta\tilde{C}_{\mu}=\tilde{A}_{\mu}-\theta c\tilde{A}_{\mu}.
\end{align}
From the construction of the dual-symmetrized Lagrangian
density $\tilde{\mathcal{L}}_{\rm QED}$, we know that the act integral will
keep invariant under a dual transformation. The corresponding Noether current reads
\begin{equation}
f_{\mu}=\frac{1}{c}\frac{\partial\tilde{\mathcal{L}}_{\rm QED}}{\partial(\partial^{\mu}\tilde{A}_{\nu})}\tilde{C}_{\nu}-c\frac{\partial\tilde{\mathcal{L}}_{\rm QED}}{\partial(\partial^{\mu}\tilde{C}_{\nu})}\tilde{A}_{\nu}.\label{eq:dual_current}
\end{equation}
Thus, the conserved quantity is given by
\begin{align}
\Lambda_M =& \frac{1}{c}\!\int \!\!d^{3}xf_{0}= -\varepsilon_{0}\!\int\!\! d^{3}x\left[\left(\partial_{0}\tilde{A}^{\nu}\right)\tilde{C}_{\nu}-\left(\partial_{0}\tilde{C}^{\nu}\right)\tilde{A}_{\nu}\right].
\end{align}
Using the subsidiary condition (\ref{eq:Subsidary}), we obtain the identity $\Lambda_M=0$. Thus, the duality symmetry does not give any new conservation law. From this point of view, the duality symmetry is trivial for QED. We can also treat $\tilde{A}_{\mu}$ and $\tilde{C}_{\mu}$ as two independent gauge fields without the subsidiary condition (\ref{eq:Subsidary}). However, in this case, no interaction between electric and magnetic charges can be obtained as shown in Appendix~\ref{sec:Hcharge2}.  

The continuity equation $\partial_{\mu}f^{\mu}=0$ will not lead to the local conservation relation between the photon helicity and spin densities as given in previous literature~\cite{barnett2012duplex,cameron2012electric,bliokh2013dual}, because the current in (\ref{eq:dual_current}) also vanishes, i.e., $f_{\mu}=0$. Applying Noether's theorem on the SO(3) rotational symmetry, we can obtain the observable part of the photon spin from $\tilde{\mathcal{L}}_{\rm QED}$~\cite{yang2020quantum},
\begin{equation}
\tilde{\boldsymbol{S}}^{\rm obs}_M = \varepsilon_0\int d^3x \left(\tilde{\boldsymbol{E}}_{\perp}\times\tilde{\boldsymbol{A}}_{\perp}+\tilde{\boldsymbol{B}}_{\perp}\times\tilde{\boldsymbol{C}}_{\perp}\right), 
\end{equation}
which only contains the spin angular momentum of transversely polarized photons. This quantity itself is conserved in the absence of charges and it can also be obtained by performing a dual transformation on its asymmetric version $\boldsymbol{S}^{\rm obs}_M = \varepsilon_0\int d^3x \boldsymbol{E}_{\perp}\times\boldsymbol{A}_{\perp}$. 

To reveal the physical nature of the dual charge~$\tilde{q}_m$, we check the conservation of charges corresponding to the global U(1) gauge symmetry. The infinitesimal U(1) symmetry transformation can be written as
\begin{align}
x_{\mu}^{\prime}& =x_{\mu},\ \tilde{A}^{\prime}_{\mu}=\tilde{A}_{\mu},\ \tilde{C}^{\prime}_{\mu} = \tilde{C}_{\mu}, \\ \psi' & =\psi+i\epsilon\psi,\ \bar{\psi}'=\bar{\psi}-i\epsilon\bar{\psi},
\end{align}
where $\epsilon$ is a coordinate independent constant. The corresponding
Noether current reads
\begin{equation}
f_{\mu}=i\epsilon\left[\frac{\partial\tilde{\mathcal{L}}}{\partial(\partial^{\mu}\psi)}\psi-\psi^{\dagger}\frac{\partial\tilde{\mathcal{L}}}{\partial(\partial^{\mu}\psi^{\dagger})}\right]=-\epsilon\bar{\psi}\gamma_{\mu}\psi.
\end{equation}
Usually, we let $\epsilon=e$ to recover the conservation of the electric
charge $Q=-e\int d^{3}x\psi^{\dagger}\psi$.
We can introduce a new Abelian phase for the Dirac field to encompass dual charges with $\epsilon=\sqrt{\tilde{q}^2_{e}+(c\varepsilon_0\tilde{q}_{m})^2}$ . We can even introduce a completely new Dirac field to carry dual charges. Then, we will obtain two independent conserved laws corresponding to the electric and dual charges separately. However, the charge-conservation law will not help us clarify the physical nature of the conserved charges, because this conservation relation due to the global gauge symmetry does not capture the effect of the linear interaction between the Dirac and EM fields.

\section{Charge-Charge interaction and quantum Lorentz force equation}
Fermi showed that the charge-charge interaction is mediated via exchanging virtual photons~\cite{Fermi1932quantum,cohen1997photons}. This gives a fundamental different picture in understanding electromagnetic interaction compared with classical Maxwell equations. However, the underlying mechanism for the interaction between dual charges, specifically the interaction between an electric charge and a dual charge, has not been clearly given. After quantizing the gauge fields $\tilde{A}_{\mu}$ and $\tilde{C}_{\mu}$, we show that both $\tilde{q}_m-\tilde{q}_m$ and $\tilde{q}_e-\tilde{q}_m$ interactions are of Coulomb form. This indicates the fact that the dual charge $\tilde{q}_m$ is actually electric charge, not magnetic charge. In electromagnetism, there only exists one type of charge, which is named as electric charge by convention. We also derive the quantum Lorentz force equation to elucidate the discrepancy with the classical theory. Finally, we show that the interaction between an electric charge and a true magnetic charge as shown in Fig.~\ref{fig:schematic} (b) cannot be mediated by exchanging photons.  

With the standard procedures in quantum field theory~\cite{cohen1997photons}, we eliminate the scalar and longitudinally polarized photons to obtain the interaction between charges (please refer to Appendix~\ref{sec:interaction} for details),\begin{widetext}
\begin{align}
\tilde{H}_{\rm charge} &\! = \!\int\! \!d^{3}x\!\!\int \!\!d^{3}x'\frac{\left[\tilde{\rho}_{e}(\boldsymbol{x})\cos\theta+c\varepsilon_{0}\tilde{\rho}_{m}(\boldsymbol{x})\sin\theta\right]\left[\tilde{\rho}_{e}(\boldsymbol{x}')\cos\theta+c\varepsilon_{0}\tilde{\rho}_{m}(\boldsymbol{x}')\sin\theta\right]}{8\pi\varepsilon_{0}|\boldsymbol{x}-\boldsymbol{x}'|}. \label{eq:H_charge}     
\end{align}\end{widetext}
Here, we show that $\tilde{q}_e-\tilde{q}_e$, $\tilde{q}_m-\tilde{q}_m$, and $\tilde{q}_e-\tilde{q}_m$ interactions all are of Coulomb form, which can only give co-axis force between charges. Furthermore, with the dual transformation relation $\rho_e = \tilde{\rho}_e\cos\theta+c\varepsilon_0\tilde{\rho}_m\sin\theta$, this charge-charge interaction $\tilde{H}_{\rm charge}$ recovers the well-known Coulomb interaction between electric charges as expected,
\begin{equation}
H_{\rm charge}=  \int d^{3}x\int d^{3}x'\frac{\rho_{e}(\boldsymbol{x})\rho_{e}(\boldsymbol{x}')}{8\pi\varepsilon_{0}|\boldsymbol{x}-\boldsymbol{x}'|} 
\end{equation} 
This provides clues to the fact that the dual charges should be interpreted as electric charges.

We show that the the duality symmetry of QED can be perfectly explained only if we interpret $\tilde{q}_m$ as electric charge. In Appendix \ref{sec:Force}, we derive the quantum Lorentz force equation within the gauge-field framework,
\begin{equation}
\frac{d}{dt}\boldsymbol{p}_{{\rm mech}}\!=\!\tilde{q}_{e}\!\left[\tilde{\boldsymbol{E}}(\boldsymbol{x})\!+\!\boldsymbol{v}\!\times\!\tilde{\boldsymbol{B}}_{\perp}(\boldsymbol{x})\right]+c\varepsilon_0\tilde{q}_{m}\!\left[c\tilde{\boldsymbol{B}}(\boldsymbol{x})\!-\!\frac{1}{c}\boldsymbol{v}\!\times\!\tilde{\boldsymbol{E}}_{\perp}(\boldsymbol{x})\right]\!,
\end{equation}
where $\boldsymbol{p}_{{\rm mech}}$ is the mechanical momentum of a Dirac particle at $\boldsymbol{x}$. Significantly different from the classical Lorentz force in Eq. (\ref{eq:Force_C}), no longitudinal EM field can enter  $\boldsymbol{v}\times\tilde{\boldsymbol{B}}$ and $\boldsymbol{v}\times\tilde{\boldsymbol{E}}$ terms. Thus, no out-of-plane force shown in Fig.~\ref{fig:schematic} (b) can be obtained from the gauge-field theory. If we interpret the dual charge as magnetic charge, the dynamics of a moving charge in the Coulomb-like magnetic field cannot be described by the quantum Lorentz force equation. Thus, disharmony will exist between the CED theory and QED gauge-field theory. In above discussion, both the dual and electric charges are carried by the same Dirac field. However, the main results will still hold if we introduce a new Dirac field to carry dual charges. 

In the derivation of the charge-charge interaction (\ref{eq:H_charge}), we have assumed that the quantized two gauge fields are of the same type of photons. Previously, D. Singleton has interpreted the quantized gauge field $\tilde{C}_{\mu}$ as magnetic photons~\cite{singleton1996electromagnetism}, which are a completely new type of photons having not been observed in experiments. However, we note that this will not change the form of the charge-charge interaction. If electric charges only interact with regular photons and magnetic charges only interact with magnetic photons, then no interaction between an electric charge and a magnetic charge  can be mediated by photons as shown in Appendix~\ref{sec:Hcharge2}. In principle, we can build another Lagrangian density manually
\begin{equation}
\tilde{\mathcal{L}}^{\prime}_{{\rm int}}=-c\bar{\psi}\gamma_{\mu}\left[\tilde{q}_{e}(\tilde{A}^{\mu}-\tilde{C}^{\mu}/c)+\varepsilon_{0}q_{m}(\tilde{C}^{\mu}+c\tilde{A}^{\mu})\right]\psi,
\end{equation}
which is also invariant under a dual transformation. However, $\tilde{\mathcal{L}}^{\prime}_{{\rm int}}$ is not a scalar and it will cause serious problems. The obtained motion equations from the Euler-Lagrange equation are not Maxwell equations as shown in Appendix~~\ref{sec:Lprime}. After eliminating the scalar photons, only a Coulomb-type interaction between the same type of charges is obtained and no $\tilde{q}_e-\tilde{q}_m$ interaction has been obtained.

We now show that true magnetic charges are not compatible with the gauge-field theory for QED. The out-of-plane force in Fig.~\ref{fig:schematic}(b) cannot be mediated by exchanging gauge photons due to the symmetry of the system. We emphasize that photon-induced interaction between two static charges must be of a central potential form because photons emitted by a static charge are spherically symmetric. Thus, the corresponding force must be along the co-axis. Similarly, the force between a static charge at $\boldsymbol{r}$ and a moving charge at $\boldsymbol{r}'$ with velocity $\boldsymbol{v}$ must be in the co-plane formed by $\boldsymbol{r}-\boldsymbol{r}'$ and $\boldsymbol{v}$. This is different from the interaction between a moving electric charge and a static magnetic dipole, where the direction of the magnetic dipole itself breaks the spherical symmetry~\cite{CHOI2004AB}. On the other hand, the central force between a static magnetic charge and an unmoving electric charge should also exist if they can exchange gauge photons. This marks a significant departure from the classical case, where the static magnetic field generated by a fixed magnetic charge will not exert any Lorentz force on an electric charge at rest. 

\section{Summary}
The magnetic charge is perfectly compatible with Maxwell's theory for CED. The electric-magnetic duality symmetry of Maxwell equations also suggests its existence. The observation of a particle carrying a magnetic charge would have a huge effect on physics. However, there are still many fundamental aspects of magnetic charges having not been understood. In this work, we unveil the mysterious mask of magnetic charges by checking the duality symmetry of QED. We show that the duality symmetry of QED, as well as CED, can be restored by interpreting the dual charges as electric charges without involving magnetic charges. More importantly, we show that true magnetic charges cannot be embedded in the gauge-field theory of QED.

\section*{Acknowledgments}
L.P.Y. thanks Prof. Zubin Jacob for helpful discussion. L.P.Y is funded by National Key R\&D Program of China (No. 2021YFE0193500). D.X. is supported by NSFC Grant No.12075025.

\appendix

\section{Dual transformation of QED Lagrangian\label{sec:DualLagrangian}}
In this section, we show how to obtain the QED Lagrangian in the symmetric representation via a dual transformation,
\begin{align}
\tilde{\boldsymbol{E}} & =\boldsymbol{E}\cos\theta-c\boldsymbol{B}\sin\theta,\ \tilde{\boldsymbol{B}}=\boldsymbol{B}\cos\theta+(\boldsymbol{E}/c)\sin\theta,\label{eq:Transform1-1}\\
\tilde{\rho}_{e} \!&\! =\!\rho_{e}\cos\theta\!-\!c\varepsilon_{0}\rho_{m}\sin\theta,\ \tilde{\rho}_{m}\!=\!\rho_{m}\cos\theta\!+\!(\rho_{e}/c\varepsilon_{0})\!\sin\theta,\label{eq:Transform2-1}\\
\tilde{A}^{\mu} \!& \!=A^{\mu}\cos\theta\!-\!(C^{\mu}/c)\sin\theta,\ \tilde{C}^{\mu}\!=\!C^{\mu}\cos\theta\!+\!cA^{\mu}\sin\theta.\label{eq:Transform3-1}
\end{align}
We note that according to the Maxwell equation, the electric filed
(a vector) changes its sign but the magnetic field (a pseudo-vector)
does not under the space inversion. Thus, the rotation angle $\theta$
must be a pseudo-scalar, which changes its sign ($\theta\rightarrow -\theta$) under the parity inversion
and the time reversal, but keep invariant under the charge reversal~\cite{jackson1999classical}. We can also verify that the electric charge $\tilde{q}_e$ is a scalar and the dual charge $\tilde{q}_m$ is a pseudo-scalar.

In the conventional asymmetric representation, the QED Lagrange density contains three part: $\mathcal{L}_{\rm QED}=\mathcal{L}_{D}+\mathcal{L}_{M}+\mathcal{L}_{{\rm int}}$.
The Lagrangina density of the Dirac field is given by 
\begin{equation}
\mathcal{L}_{D}=i\hbar c\bar{\psi}\gamma^{\mu}\partial_{\mu}\psi-mc^{2}\bar{\psi}\psi,
\end{equation}
where $\bar{\psi}=\psi^{\dagger}\gamma^{0}$ and
\begin{equation}
\gamma^{0}=\beta,\ \gamma^{i}=\beta\alpha_{i},\ i=1,2,3
\end{equation}
with
\begin{equation}
\beta=\left[\begin{array}{cc}
I & 0\\
0 & -I
\end{array}\right],\ \alpha_{i}=\left[\begin{array}{cc}
0 & \sigma_{i}\\
\sigma_{i} & 0
\end{array}\right],
\end{equation}
the $2\times2$ identity matrix $I$, and the Pauli matrices $\sigma_{i}$.
To quantize the Maxwell field covariantly, we take the Fermi Lagrange density instead of the standard one for the EM field~\cite{greiner2013field,cohen1997photons},
\begin{equation}
\mathcal{L}_{M}=-\frac{1}{2\mu_{0}}(\partial_{\mu}A^{\nu})(\partial^{\mu}A_{\nu}),
\end{equation}
and the Lorenz
gauge $\partial^{\mu}A_{\mu}=0$. The interaction between Dirac-Maxwell fields is given by
\begin{equation}
\mathcal{L}_{{\rm int}}=-q_{e}c\bar{\psi}\gamma_{\mu}A^{\mu}\psi.
\end{equation}

Now we show how to obtain the dual symmetric Lagrangian $\tilde{\mathcal{L}}_{\rm QED} = \mathcal{L}_D+\tilde{\mathcal{L}}_M+\tilde{\mathcal{L}}_{\rm int}$ from asymmetric one $\mathcal{L}_{\rm QED}$. The Dirac Lagrangian density does not change under a dual transformation. The Fermi Lagrangian density for the EM fields changes into
\begin{align}
\tilde{\mathcal{L}}_{M} & =-\frac{1}{2\mu_{0}}\left\{ (\partial_{\mu}\tilde{A}^{\nu})(\partial^{\mu}\tilde{A}_{\nu})\cos^{2}\theta+\frac{1}{c^{2}}(\partial_{\mu}\tilde{C}^{\nu})(\partial^{\mu}\tilde{C}_{\nu})\sin^{2}\theta\right. \nonumber\\
& \left.+\frac{1}{c}\left[(\partial_{\mu}\tilde{A}^{\nu})(\partial^{\mu}\tilde{C}_{\nu})+(\partial_{\mu}\tilde{C}^{\nu})(\partial^{\mu}\tilde{A}_{\nu})\right]\sin\theta\cos\theta\right\} \\
 & =-\frac{1}{2\mu_{0}}\left\{ (\partial_{\mu}\tilde{A}^{\nu})(\partial^{\mu}\tilde{A}_{\nu})+\frac{1}{c^{2}}(\partial_{\mu}\tilde{C}^{\nu})(\partial^{\mu}\tilde{C}_{\nu})\right\}  \\
 & =-\frac{1}{2}\left\{ \frac{1}{\mu_{0}}(\partial_{\mu}\tilde{A}^{\nu})(\partial^{\mu}\tilde{A}_{\nu})+\varepsilon_{0}(\partial_{\mu}\tilde{C}^{\nu})(\partial^{\mu}\tilde{C}_{\nu})\right\}.
\end{align}
In the second step, we have used the subsidiary condition  $\tilde{q}_{e}\tilde{C}^{\mu}-\tilde{q}_{m}\tilde{A}^{\mu}/\mu_{0}=0$
(i.e., $C^{\mu}=\tilde{C}^{\mu}\cos\theta-c\tilde{A}^{\mu}\sin\theta=0$). The interaction part now reads
\begin{align}
\mathcal{\tilde{L}}_{{\rm int}} \!& =\!-\left(\tilde{q}_{e}\cos\theta\!+\!c\varepsilon_{0}\tilde{q}_{m}\sin\theta\right)c\bar{\psi}\gamma_{\mu}\left(\tilde{A}^{\mu}\cos\theta\!+\!(\tilde{C}^{\mu}/c)\sin\theta\right)\psi\nonumber\\
 & =-c\bar{\psi}\gamma_{\mu}(\tilde{q}_{e}\tilde{A}^{\mu}+\varepsilon_{0}\tilde{q}_{m}\tilde{C}^{\mu})\psi.
\end{align}

We can also check that the two EM tensors also transform like a vector in the dual space,
\begin{equation}
F^{\mu\nu}  =  \tilde{F}^{\mu\nu}\cos\theta + (1/c)\tilde{G}^{\mu\nu}\sin\theta,\  G^{\mu\nu} = \tilde{G}^{\mu\nu}\cos\theta - c\tilde{F}^{\mu\nu}\sin\theta.
\end{equation}
After a dual transformation, the stand Lagrangian density of light $\mathcal{L}_{\rm M,st}=-F^{\mu\nu}F_{\mu\nu}/4\mu_0$ changes into
\begin{align}
\mathcal{\tilde{L}}_{{\rm M,st}} & =-\frac{1}{4\mu_0}\left(\tilde{F}^{\mu\nu}\cos\theta\!+\!\frac{1}{c}\tilde{G}^{\mu\nu}\sin\theta\right)\left(\tilde{F}_{\mu\nu}\cos\theta\!+\!\frac{1}{c}\tilde{G}_{\mu\nu}\sin\theta\right)\nonumber\\
 & =-\frac{1}{4\mu_0}\tilde{F}^{\mu\nu}\tilde{F}_{\mu\nu}-\frac{\varepsilon_0}{4}\tilde{G}^{\mu\nu}\tilde{G}_{\mu\nu},
\end{align}
where the subsidiary condition $\tilde{G}^{\mu\nu}\cos\theta = c\tilde{F}^{\mu\nu}\sin\theta$ has been used. Here, we emphasize that the Lagrangian density $\mathcal{\tilde{L}}_{{\rm M,st}}$ is significantly different from the one introduced in Ref.~\cite{cameron2012electric,bliokh2013dual}, in which the dual EM tensor $\mathcal{F}^{\mu\nu}=\epsilon^{\mu\nu\alpha\beta}F_{\alpha\beta}/2$ part has been added manually. As explained in the main text, our defined EM tensor $\tilde{G}^{\mu\nu}$ is not the dual EM tensor.

\section{From Dual-symmetric Lagrangian to Dual-symmetric Maxwell equations \label{sec:MaxwellEqs}}
In this section, we show the electric-magnetic duality symmetry embedded in the Lagrangian~$\tilde{\mathcal{L}}_{\rm QED}$ given in Appendix~\ref{sec:DualLagrangian}. Via the Euler-Lagrange equation, we re-derive the symmetric Maxwell equations in Eq.~(\ref{eq:Maxwell}) without applying the subsidiary condition.

The Euler-Lagrange equations for the gauge potentials $\tilde{A}_{\mu}$ and $\tilde{C}_{\mu}$
are given by
\begin{align}
\partial_{\nu}\partial^{\nu}\tilde{A}^{\mu} & =\mu_{0}\tilde{q}_{e}c\bar{\psi}\gamma^{\mu}\psi,\\
\partial_{\nu}\partial^{\nu}\tilde{C}^{\mu} & =\tilde{q}_{m}c\bar{\psi}\gamma^{\mu}\psi.
\end{align}
The relation between the EM fields and the gauge potentials are given in Eqs.~(\ref{eq:E}) and (\ref{eq:B}). The divergence of $\tilde{\boldsymbol{E}}$ and $\tilde{\boldsymbol{B}}$ gives
\begin{align}
\boldsymbol{\nabla}\cdot\tilde{\boldsymbol{E}} & =-c\left[\partial^{0}\boldsymbol{\nabla}\cdot\tilde{\boldsymbol{A}}+\nabla^{2}\tilde{A}^{0}\right],\\
\boldsymbol{\nabla}\cdot\tilde{\boldsymbol{B}} & =-\frac{1}{c}\left[\partial^{0}\boldsymbol{\nabla}\cdot\tilde{\boldsymbol{C}}+\nabla^{2}\tilde{C}^{0}\right].
\end{align}
Using the motion equations
\begin{align}
\left(\partial_{0}\partial^{0}-\nabla^{2}\right)\tilde{A}^{0} & =\mu_{0}c\tilde{\rho}_{e},\\
\left(\partial_{0}\partial^{0}-\nabla^{2}\right)\tilde{C}^{0} & =c\tilde{\rho}_{m},
\end{align}
with $\tilde{\rho}_{e}=q_{e}\psi^{\dagger}\psi$ and $\tilde{\rho}_{m}=q_{m}\psi^{\dagger}\psi$,
we have
\begin{align}
\boldsymbol{\nabla}\cdot\tilde{\boldsymbol{E}} & =-c\left[\partial^{0}\boldsymbol{\nabla}\cdot\tilde{\boldsymbol{A}}+\partial_{0}\partial^{0}\tilde{A}^{0}-\mu_{0}c\tilde{\rho}_{e}\right]=\tilde{\rho}_{e}/\varepsilon_{0},\\
\boldsymbol{\nabla}\cdot\tilde{\boldsymbol{B}} & =-\frac{1}{c}\left[\partial^{0}\boldsymbol{\nabla}\cdot\tilde{\boldsymbol{C}}+\partial_{0}\partial^{0}\tilde{C}^{0}-c\tilde{\rho}_{m}\right] =\tilde{\rho}_{m},
\end{align}
where we have used the Lorenz gauge condition for the two potentials $\partial_{0}\tilde{A}^{0}+\boldsymbol{\nabla}\cdot\tilde{\boldsymbol{A}}=\partial_{0}\tilde{C}^{0}+\boldsymbol{\nabla}\cdot\tilde{\boldsymbol{C}}=0$. 

The curl of $\tilde{\boldsymbol{E}}$ and $\tilde{\boldsymbol{B}}$ gives
\begin{align}
\boldsymbol{\nabla}\times\tilde{\boldsymbol{E}} & =-(c\partial^{0}\boldsymbol{\nabla}\times\tilde{\boldsymbol{A}}+\boldsymbol{\nabla}\times\boldsymbol{\nabla}\times\tilde{\boldsymbol{C}})\nonumber\\
 & =-\left[c\partial^{0}\boldsymbol{\nabla}\times\tilde{\boldsymbol{A}}+\boldsymbol{\nabla}(\boldsymbol{\nabla}\cdot\tilde{\boldsymbol{C}})-\nabla^{2}\tilde{\boldsymbol{C}}\right],\\
\boldsymbol{\nabla}\times\tilde{\boldsymbol{B}} & =-\left[(1/c)\partial^{0}\boldsymbol{\nabla}\times\tilde{\boldsymbol{C}}-\boldsymbol{\nabla}\times\boldsymbol{\nabla}\times\tilde{\boldsymbol{A}}\right]\nonumber\\
 & =-\left[(1/c)\partial^{0}\boldsymbol{\nabla}\times\tilde{\boldsymbol{C}}-\boldsymbol{\nabla}(\boldsymbol{\nabla}\cdot\tilde{\boldsymbol{A}})+\nabla^{2}\tilde{\boldsymbol{A}}\right].
\end{align}
Using the motion equations
\begin{align}
\left(\partial_{0}\partial^{0}-\nabla^{2}\right)\tilde{\boldsymbol{A}} & =\mu_{0}\tilde{\boldsymbol{j}}_{e},\\
\left(\partial_{0}\partial^{0}-\nabla^{2}\right)\tilde{\boldsymbol{C}} & =\tilde{\boldsymbol{j}}_{m},
\end{align}
with $\tilde{\boldsymbol{j}}_{e}=q_{e}\psi^{\dagger}c\boldsymbol{\alpha}\psi$
and $\tilde{\boldsymbol{j}}_{m}=q_{m}\psi^{\dagger}c\boldsymbol{\alpha}\psi$,
we have
\begin{align}
\boldsymbol{\nabla}\times\tilde{\boldsymbol{E}} & =-\left[c\partial^{0}\boldsymbol{\nabla}\times\tilde{\boldsymbol{A}}+\boldsymbol{\nabla}(\boldsymbol{\nabla}\cdot\tilde{\boldsymbol{C}})-\partial_{0}\partial^{0}\tilde{\boldsymbol{C}}+\tilde{\boldsymbol{j}}_{m}\right]\\
 & =\partial^{0}\left[-c\boldsymbol{\nabla}\times\tilde{\boldsymbol{A}}+\boldsymbol{\nabla}\tilde{C}^{0}+\partial_{0}\tilde{\boldsymbol{C}}\right]-\tilde{\boldsymbol{j}}_{m}\\
 & =-\frac{\partial}{\partial t}\tilde{\boldsymbol{B}}-\tilde{\boldsymbol{j}}_{m},
\end{align}
and 
\begin{align*}
\boldsymbol{\nabla}\times\tilde{\boldsymbol{B}} & =-\left[(1/c)\partial^{0}\boldsymbol{\nabla}\times\tilde{\boldsymbol{C}}-\boldsymbol{\nabla}(\boldsymbol{\nabla}\cdot\tilde{\boldsymbol{A}})+\partial_{0}\partial^{0}\tilde{\boldsymbol{A}}-\mu_{0}\tilde{\boldsymbol{j}}_{e}\right]\\
 & =-\frac{1}{c}\partial^{0}\left[\boldsymbol{\nabla}\times\tilde{\boldsymbol{C}}+c\boldsymbol{\nabla}\tilde{A}^{0}+c\partial_{0}\tilde{\boldsymbol{A}}\right]+\mu_{0}\tilde{\boldsymbol{j}}_{e}\\
 & =\frac{1}{c^{2}}\frac{\partial}{\partial t}\tilde{\boldsymbol{E}}+\mu_{0}\tilde{\boldsymbol{j}}_{e},
\end{align*}
i.e.,
\begin{equation}
\boldsymbol{\nabla}\times\tilde{\boldsymbol{H}}=\frac{\partial}{\partial t}\tilde{\boldsymbol{D}}+\tilde{\boldsymbol{j}}_{e}.
\end{equation}
Here, we see that the dual-symmetric Maxwell equations can be derived from the dual-symmetric Lagrangian without the subsidiary condition. The two charges $q_e$ and $q_m$ correspond to the electric and magnetic charges in CED, respectively.

We note that in CED, the subsidiary is only required to transform to the symmetry representation from the asymmetric one. If we start from the symmetric representation, the magnetic charge is completely compatible with the symmetry Maxwell equations. The subsidiary condition does not play an essential role in electromagnetic interaction. This is not true in QED as shown in Appendix~\ref{sec:interaction}.

\section{QED with one gauge field \label{sec:interaction}}
In this section, we show the quantization of the fields in the symmetric representation. We give the standard canonical quantization recipe in the asymmetric representation first. The canonical momentum of light is defined as 
\begin{equation}
\pi_{\mu}=\frac{\partial\mathcal{L}}{\partial(\partial_{0}A^{\mu})}=-\frac{1}{\mu_{0}}\partial^{0}A_{\mu}.
\end{equation}
The quantization of both the Maxwell and Dirac fields will be realized by postulating the following equal-time commutation relations
\begin{align}
[\psi_{r}(\boldsymbol{x}),\psi_{r'}^{\dagger}(\boldsymbol{x}')]_{+} & =\delta_{rr'}\delta^{3}(\boldsymbol{x}-\boldsymbol{x}'),\\{}
[\psi_{r}(\boldsymbol{x}),\psi_{r'}(\boldsymbol{x}')]_{+} & =[\psi_{r}^{\dagger}(\boldsymbol{x}),\psi_{r'}^{\dagger}(\boldsymbol{x}')]_{+}=0,
\end{align}
and
\begin{align}
[A^{\mu}(\boldsymbol{x},t),\pi^{\nu}(\boldsymbol{x}',t)] & =i\hbar cg^{\mu\nu}\delta^{3}(\boldsymbol{x}-\boldsymbol{x}'),\\{}
[A^{\mu}(\boldsymbol{x},t),A^{\nu}(\boldsymbol{x}',t)] & =[\pi^{\mu}(\boldsymbol{x},t),\pi^{\nu}(\boldsymbol{x}',t)]=0.
\end{align}
The operators of Maxwell and Dirac fields  commute with each other. The total QED Hamiltonian can be split into three parts $
H_{\rm QED}=H_{D}+H_{M}+H_{{\rm int}}$~\cite{yang2020quantum,cohen1997photons}, where the Hamiltonian for the Dirac field, Maxwell field, and their interaction are given by
\begin{align}
H_{D} & =\int d^3x\psi^{\dagger}\left[-i\hbar c\boldsymbol{\alpha}\cdot\boldsymbol{\nabla}+mc^{2}\beta\right]\psi,\\
H_{M} & =-\int d^3x\frac{1}{2\mu_{0}}\left[(\partial^{0}A^{\sigma})(\partial^{0}A_{\sigma})+(\boldsymbol{\nabla}A^{\sigma})\cdot(\boldsymbol{\nabla}A_{\sigma})\right],\\
H_{{\rm int}} & =\int d^3xc\bar{\psi}\gamma_{\mu}qA^{\mu}\psi.
\end{align}

In the symmetric representation, we have two dependent canonical momenta for the Maxwell field
\begin{align}
\tilde{\pi}_{A}^{\mu} & =\frac{\partial\tilde{\mathcal{L}}}{\partial(\partial^{0}\tilde{A}_{\mu})}=-\frac{1}{\mu_{0}}\partial_{0}\tilde{A}^{\mu}.\label{eq:Pi_A}\\
\tilde{\pi}_{C}^{\mu} & =\frac{\partial\tilde{\mathcal{L}}}{\partial(\partial^{0}\tilde{C}_{\mu})}=-\varepsilon_{0}\partial_{0}\tilde{C}^{\mu},\label{eq:Pi_C}
\end{align}
Using the subsidiary condition (\ref{eq:Subsidary}), we have $\tilde{\pi}_{A}^{\mu}=\pi^{\mu}\cos\theta$ and $\tilde{\pi}_{C}^{\mu}=(1/c)\pi^{\mu}\sin\theta$. 
The quantization conditions for the Dirac field remain the same, but the equal-time commutation relations for the Maxwell field now change into
\begin{align}
[\tilde{A}^{\mu}(\boldsymbol{x},t),\tilde{\pi}_{A}^{\nu}(\boldsymbol{x}',t)] & =i\hbar cg^{\mu\nu}\cos^{2}\theta\delta^{3}(\boldsymbol{x}-\boldsymbol{x}'),\\{}
[\tilde{C}^{\mu}(\boldsymbol{x},t),\tilde{\pi}_{C}^{\nu}(\boldsymbol{x}',t)] & =i\hbar cg^{\mu\nu}\sin^{2}\theta\delta^{3}(\boldsymbol{x}-\boldsymbol{x}'),\\{}
[\tilde{A}^{\mu}(\boldsymbol{x},t),\tilde{\pi}_{C}^{\nu}(\boldsymbol{x}',t)] & =i\hbar g^{\mu\nu}\sin\theta\cos\theta\delta^{3}(\boldsymbol{x}-\boldsymbol{x}'),\\{}
[\tilde{C}^{\mu}(\boldsymbol{x},t),\tilde{\pi}_{A}^{\nu}(\boldsymbol{x}',t)] & =i\hbar c^{2}g^{\mu\nu}\sin\theta\cos\theta\delta^{3}(\boldsymbol{x}-\boldsymbol{x}'),\\
[\tilde{A}^{\mu}(\boldsymbol{x},t),\tilde{A}^{\nu}(\boldsymbol{x}',t)] & =[\tilde{\pi}_{A}^{\mu}(\boldsymbol{x},t),\tilde{\pi}_{A}^{\nu}(\boldsymbol{x}',t)]=0,\\{}
[\tilde{C}^{\mu}(\boldsymbol{x},t),\tilde{C}^{\nu}(\boldsymbol{x}',t)] & =[\tilde{\pi}_{C}^{\mu}(\boldsymbol{x},t),\tilde{\pi}_{C}^{\nu}(\boldsymbol{x}',t)]=0.
\end{align}
The QED Hamiltonian in the symmetric representation is given by $\tilde{H}_{\rm QED}=\tilde{H}_{D}+\tilde{H}_{M}+\tilde{H}_{{\rm int}}$
with
\begin{align}
\tilde{H}_{D} = & H_{D} =\int d^3x\psi^{\dagger}\left[-i\hbar c\boldsymbol{\alpha}\cdot\boldsymbol{\nabla}+mc^{2}\beta\right]\psi,\\
\tilde{H}_{M} = &-\frac{1}{2\mu_{0}}\int d^{3}x\left[(\partial^{0}\tilde{A}^{\sigma})(\partial^{0}\tilde{A}_{\sigma})+(\boldsymbol{\nabla}\tilde{A}^{\sigma})\cdot(\boldsymbol{\nabla}\tilde{A}_{\sigma})\right]\nonumber \\
&-\int d^{3}x\frac{\varepsilon_{0}}{2}\left[(\partial^{0}\tilde{C}^{\sigma})(\partial^{0}\tilde{C}_{\sigma})+(\boldsymbol{\nabla}\tilde{C}^{\sigma})\cdot(\boldsymbol{\nabla}\tilde{C}_{\sigma})\right],\\
\tilde{H}_{{\rm int}} = & \int d^{3}xc\bar{\psi}\gamma_{\mu}(\tilde{q}_{e}\tilde{A}^{\mu}+\varepsilon_{0}\tilde{q}_{m}\tilde{C}^{\mu})\psi.
\end{align}

Next, we will use the quantum Lorenz gauge to derive the interaction between the charges. We give the plane-wave expansion of the four Maxwell-field operators
\begin{equation}
\tilde{A}^{\mu} \! =\!\!\int\!\! d^{3}k\!\sum_{\lambda=0}^{3}\!\!\sqrt{\frac{\hbar}{2\varepsilon_{0}\omega_{\boldsymbol{k}}(2\pi)^{3}}}\left[a_{\boldsymbol{k},\lambda}\epsilon^{\mu}(\boldsymbol{k},\lambda)e^{i\boldsymbol{k}\cdot\boldsymbol{x}}\!+\!{\rm h.c.}\right]\cos\theta,
\end{equation}
\begin{equation}
\tilde{C}^{\mu} \! =\!\!\int\!\! d^{3}k\!\sum_{\lambda=0}^{3}\sqrt{\frac{c^{2}\hbar}{2\varepsilon_{0}\omega_{\boldsymbol{k}}(2\pi)^{3}}}\left[a_{\boldsymbol{k},\lambda}\epsilon^{\mu}(\boldsymbol{k},\lambda)e^{i\boldsymbol{k}\cdot\boldsymbol{x}}+{\rm h.c.}\right]\sin\theta,
\end{equation}
\begin{equation}
\tilde{\pi}_{A}^{\mu} \! =\!\!\int\!\! d^{3}k\!\!\sum_{\lambda=0}^{3}\!\!\sqrt{\frac{\hbar\omega_{\boldsymbol{k}}}{2\mu_{0}(2\pi)^{3}}}\left[a_{\boldsymbol{k},\lambda}\epsilon^{\mu}(\boldsymbol{k},\lambda)e^{i\boldsymbol{k}\cdot\boldsymbol{x}}\!-\!{\rm h.c.}\right]\cos\theta,
\end{equation}
\begin{equation}
\tilde{\pi}_{C}^{\mu} \! =\!\!\int\!\! d^{3}k\!\sum_{\lambda=0}^{3}\sqrt{\frac{\varepsilon_{0}\hbar\omega_{\boldsymbol{k}}}{2(2\pi)^{3}}}\left[a_{\boldsymbol{k},\lambda}\epsilon^{\mu}(\boldsymbol{k},\lambda)e^{i\boldsymbol{k}\cdot\boldsymbol{x}}-{\rm h.c.}\right]\sin\theta,
\end{equation}
where $\omega_{\boldsymbol{k}}=c|\boldsymbol{k}|$ is frequency of the mode with wave vector $\boldsymbol{k}$ and the unit vectors $\epsilon(\boldsymbol{k},\lambda)$ describe the four polarization photons. Following the convention~\cite{greiner2013field,cohen1997photons}, we let the two unit vectors $\epsilon(\boldsymbol{k},1)$ and $\epsilon(\boldsymbol{k},2)$ denote the two transverse modes, $\epsilon(\boldsymbol{k},3)=(0,\boldsymbol{k}/|\boldsymbol{k}|)$ for the longitudinal photon, and $\epsilon(\boldsymbol{k},0)=(1,0,0,0)$ for the scalar photon. In the following, we also use  $\boldsymbol{\epsilon}(\boldsymbol{k},\lambda)$ to denote the spatial part of the four-vector $\epsilon(\boldsymbol{k},\lambda)$. The ladder operators satisfy the bosonic commutation relations $
[a_{\boldsymbol{k},\lambda},a_{\boldsymbol{k}',\lambda'}^{\dagger}]  =-g_{\lambda\lambda'}\delta^{3}(\boldsymbol{k}-\boldsymbol{k}')$ and $  [a_{\boldsymbol{k},\lambda},a_{\boldsymbol{k}',\lambda'}]  = [a_{\boldsymbol{k},\lambda}^{\dagger},a_{\boldsymbol{k}',\lambda'}^{\dagger}]=0$. 

Now, we can split the Hamiltonian of the Maxwell field into three parts
\begin{align}
\tilde{H}_M = & \tilde{H}_{M}^{T}+\tilde{H}_{M}^{L}+\tilde{H}_{M}^{S} \\
= &\int d^{3}k\hbar\omega_{\boldsymbol{k}}(a_{\boldsymbol{k},1}^{\dagger}a_{\boldsymbol{k},1}+a_{\boldsymbol{k},2}^{\dagger}a_{\boldsymbol{k},2}) \nonumber\\
& + \int d^{3}k\hbar\omega_{\boldsymbol{k}}a_{\boldsymbol{k},3}^{\dagger}a_{\boldsymbol{k},3}-\int d^{3}k\hbar\omega_{\boldsymbol{k}}a_{\boldsymbol{k},0}^{\dagger}a_{\boldsymbol{k},0}, 
\end{align}
where the first term describes the transversely polarized photons, the third term is for longitudinally polarized photons, and the last term denote the scalar photons with negative frequencies. After plane-wave expansion, we see that the Hamiltonian of photons $\tilde {H}_M$ in the symmetric representation reduces to its asymmetric counterpart exactly~\cite{cohen1997photons}. Similarly, the negative frequency problem will be solved via the quantum Lorenz gauge condition as shown in the following.

Using the definition of the charge density and current,
\begin{align}
\tilde{\rho}_{e}(\boldsymbol{x}) & =\tilde{q}_{e}\psi^{\dagger}(\boldsymbol{x})\psi(\boldsymbol{x}),\ \tilde{\rho}_{m}=\tilde{q}_{m}\psi^{\dagger}(\boldsymbol{x})\psi(\boldsymbol{x}),\\
\tilde{\boldsymbol{j}}_{e}(\boldsymbol{x}) & =\tilde{q}_{e}c\psi^{\dagger}(\boldsymbol{x})\boldsymbol{\alpha}\psi(\boldsymbol{x}),\ \tilde{\boldsymbol{j}}_{m}=\tilde{q}_{m}c\psi^{\dagger}(\boldsymbol{x})\boldsymbol{\alpha}\psi(\boldsymbol{x}),
\end{align}
the interaction parts can be expressed as $\tilde{H}_{\rm int}=\tilde{H}_{{\rm int}}^{T}+\tilde{H}_{{\rm int}}^{L}+\tilde{H}_{{\rm int}}^{S}$ with\begin{widetext}
\begin{align}
\tilde{H}_{{\rm int}}^{T}\!+\!\tilde{H}_{{\rm int}}^{L}
= -\int d^{3}x\left[\tilde{\boldsymbol{j}}_{e}(\boldsymbol{x})\cdot\tilde{\boldsymbol{A}}(\boldsymbol{x})+\varepsilon_{0}\tilde{\boldsymbol{j}}_{m}(\boldsymbol{x})\cdot\tilde{\boldsymbol{C}}(\boldsymbol{x})\right]
= -\int d^{3}k\hbar\omega_{\boldsymbol{k}}\sum_{\lambda=1}^{3}\left\{ a_{\boldsymbol{k},\lambda}^{\dagger}\left[\boldsymbol{\xi}_{e}(\boldsymbol{k})\cos\theta+\boldsymbol{\xi}_{m}(\boldsymbol{k})\sin\theta\right]\cdot\boldsymbol{\epsilon}(\boldsymbol{k},\lambda)+{\rm h.c.}\right\} ,
\end{align}\end{widetext}
\begin{align}
\tilde{H}_{{\rm int}}^{S} & =c\int d^{3}x\left[\tilde{\rho}_{e}(\boldsymbol{x})\tilde{A}_{0}(\boldsymbol{x})+\varepsilon_{0}\tilde{\rho}_{m}\tilde{C}_{0}(\boldsymbol{x})\right]\\
& =\!\!\!\int\!\! d^{3}k\hbar\omega_{\boldsymbol{k}}\!\left\{ \left[\xi_{e,0}(\boldsymbol{k})\cos\theta\!+\!\xi_{m,0}(\boldsymbol{k})\sin\theta\right]a_{\boldsymbol{k},0}^{\dagger}\!+\!{\rm h.c.}\right\}\!.
\end{align}
where we have defined the following quantities
\begin{equation}
\boldsymbol{\xi}_{e}(\boldsymbol{k})  =\frac{1}{\hbar\omega_{\boldsymbol{k}}}\sqrt{\frac{\hbar}{2\varepsilon_{0}\omega_{\boldsymbol{k}}(2\pi)^{3}}}\int d^{3}x\tilde{\boldsymbol{j}}_{e}(\boldsymbol{x})e^{-i\boldsymbol{k}\cdot\boldsymbol{x}},
\end{equation}
\begin{equation}
\boldsymbol{\xi}_{m}(\boldsymbol{k})  =\frac{c\varepsilon_{0}}{\hbar\omega_{\boldsymbol{k}}}\sqrt{\frac{\hbar}{2\varepsilon_{0}\omega_{\boldsymbol{k}}(2\pi)^{3}}}\int d^{3}x\tilde{\boldsymbol{j}}_{m}(\boldsymbol{x})e^{-i\boldsymbol{k}\cdot\boldsymbol{x}},
\end{equation}
\begin{equation}
\xi_{e,0}(\boldsymbol{k})  =\frac{c}{\hbar\omega_{\boldsymbol{k}}}\sqrt{\frac{\hbar}{2\varepsilon_{0}\omega_{\boldsymbol{k}}(2\pi)^{3}}}\int d^{3}x\tilde{\rho}_{e}(\boldsymbol{x})e^{-i\boldsymbol{k}\cdot\boldsymbol{x}},
\end{equation}
\begin{equation}
\xi_{m,0}(\boldsymbol{k}) =\frac{c^{2}\varepsilon_{0}}{\hbar\omega_{\boldsymbol{k}}}\sqrt{\frac{\hbar}{2\varepsilon_{0}\omega_{\boldsymbol{k}}(2\pi)^{3}}}\int d^{3}x\tilde{\rho}_{m}(\boldsymbol{x})e^{-i\boldsymbol{k}\cdot\boldsymbol{x}}.
\end{equation}

To eliminate the scalar photons,
we evaluate the Gupta-Bleuler condition, i.e., the quantum version fo the Lorenz gauge condition, $\partial^{\mu}\tilde{A}_{\mu}^{(+)}\left|\Phi\right\rangle =\partial^{\mu}\tilde{C}_{\mu}^{(+)}\left|\Phi\right\rangle =0$
for the Dirac-Maxwell field~\cite{cohen1997photons}. Here, $|\Phi\rangle$ is an arbitrary physical state. In the Heisenberg picture, the motion
equation of the scalar ladder operators are given by
\begin{align}
\dot{a}_{\boldsymbol{k},0} & =\frac{i}{\hbar}\left[\tilde{H}_{{\rm QED}},a_{\boldsymbol{k},0}\right]\nonumber\\
& =-i\omega_{\boldsymbol{k}}\left[a_{\boldsymbol{k},0}-\xi_{e,0}(\boldsymbol{k})\cos\theta-\xi_{m,0}(\boldsymbol{k})\sin\theta\right],\\
\dot{a}_{\boldsymbol{k},0}^{\dagger} & =\frac{i}{\hbar}\left[\tilde{H}_{{\rm QED}},a_{\boldsymbol{k},0}^{\dagger}\right]\nonumber\\
& = i\omega_{\boldsymbol{k}}\left[a_{\boldsymbol{k},0}-\xi_{e,0}^{\dagger}(\boldsymbol{k})\cos\theta-\xi_{m,0}^{\dagger}(\boldsymbol{k})\sin\theta\right].
\end{align}
The Gupta-Bleuler must hold for all plane-wave modes. This requires
\begin{equation}
\left[a_{\boldsymbol{k},3}-a_{\boldsymbol{k},0}+\xi_{e,0}(\boldsymbol{k})\cos\theta+\xi_{m,0}(\boldsymbol{k})\sin\theta\right]\left|\Phi\right\rangle =0,
\end{equation}
and
\begin{equation}
\left\langle \Phi\right|\left[a_{\boldsymbol{k},3}^{\dagger}-a_{\boldsymbol{k},0}^{\dagger}+\xi_{e,0}^{\dagger}(\boldsymbol{k})\cos\theta+\xi_{m,0}^{\dagger}(\boldsymbol{k})\sin\theta\right]=0.
\end{equation}

Now, we show that, for any physical state $\left|\Phi\right\rangle $, the mean value of $\left\langle \Phi\right|\tilde{H}_{M}^{L}+\tilde{H}_{M}^{S}+\tilde{H}_{{\rm int}}^{S}\left|\Phi\right\rangle$ gives the Coulomb interaction between charges\begin{widetext}
\begin{align}
\left\langle \Phi\right|\tilde{H}_{M}^{L}+\tilde{H}_{M}^{S}+\tilde{H}_{{\rm int}}^{S}\left|\Phi\right\rangle
= & \left\langle \Phi\right|\int d^{3}k\hbar\omega_{\boldsymbol{k}}\left\{ a_{\boldsymbol{k},3}^{\dagger}a_{\boldsymbol{k},3}-a_{\boldsymbol{k},0}^{\dagger}a_{\boldsymbol{k},0}+a_{\boldsymbol{k},0}^{\dagger}\left[\xi_{e,0}(\boldsymbol{k})\cos\theta+\xi_{m,0}(\boldsymbol{k})\sin\theta\right]\right.\nonumber\\
& \left.+[\xi_{e,0}^{\dagger}(\boldsymbol{k})\cos\theta+\xi_{m,0}^{\dagger}(\boldsymbol{k})\sin\theta]a_{\boldsymbol{k},0}\right\} \left|\Phi\right\rangle \\
= & \left\langle \Phi\right|\int d^{3}k\hbar\omega_{\boldsymbol{k}}\left[\xi_{e,0}^{\dagger}(\boldsymbol{k})\cos\theta+\xi_{m,0}^{\dagger}(\boldsymbol{k})\sin\theta\right]\left[\xi_{e,0}(\boldsymbol{k})\cos\theta+\xi_{m,0}(\boldsymbol{k})\sin\theta\right]\left|\Phi\right\rangle \\
= & \left\langle \Phi\right|\int d^{3}k\frac{\left[\tilde{\rho}_{e}^{\dagger}(\boldsymbol{k})\cos\theta+c\varepsilon_{0}\tilde{\rho}_{m}^{\dagger}(\boldsymbol{k})\sin\theta\right]\left[\tilde{\rho}_{e}(\boldsymbol{k})\cos\theta+c\varepsilon_{0}\tilde{\rho}_{m}(\boldsymbol{k})\sin\theta\right]}{2\varepsilon_{0}|\boldsymbol{k}|^{2}}\left|\Phi\right\rangle \\
= & \left\langle \Phi\right|\int d^{3}x\int d^{3}x'\frac{\left[\tilde{\rho}_{e}(\boldsymbol{x})\cos\theta+c\varepsilon_{0}\tilde{\rho}_{m}(\boldsymbol{x})\sin\theta\right]\left[\tilde{\rho}_{e}(\boldsymbol{x}')\cos\theta+c\varepsilon_{0}\tilde{\rho}_{m}(\boldsymbol{x}')\sin\theta\right]}{8\pi\varepsilon_{0}|\boldsymbol{x}-\boldsymbol{x}'|}\left|\Phi\right\rangle \\
= & \left\langle \Phi\right|\int d^{3}x\int d^{3}x'\frac{\rho_{e}(\boldsymbol{x})\rho_{e}(\boldsymbol{x}')}{8\pi\varepsilon_{0}|\boldsymbol{x}-\boldsymbol{x}'|}\left|\Phi\right\rangle 
\end{align}\end{widetext}
We note that the remaining interacting part $\tilde{H}_{\rm int}^{L}$ should be absorbed into $\tilde{H}_D$ to make the total Hamiltonian keep invariant under a gauge transformation.

\section{Quantum Lorentz force equation \label{sec:Force}}
We now derive the quantum Lorentz force equation in the first-quantization picture of the charge particles. We give the Lorentz force equation in the asymmetric representation and then perform a dual rotation to obtain the counterpart in the symmetric representation.

In the previous section, we have eliminated the scalar photons to obtain the Coulomb interaction between charges. Now, it is more convenient to derive the Lorentz force equation in the Coulomb gauge. The Hamiltonian for a multi-charge system is given by (see Complement B of Chap. V in~\cite{cohen1997photons}),
\begin{equation}
H=\sum_{l}c\boldsymbol{\alpha}^{(l)}\cdot\left[\boldsymbol{p}^{(l)}-q_{e}\boldsymbol{A}_{\perp}(\boldsymbol{x}^{(l)})\right]+H_{{\rm Charge}}+H_{M}^{T},
\end{equation}
with the Hamiltonian for the transverse photons $H_M=\int d^3k\sum_{\lambda =1,2}\hbar\omega_{\boldsymbol{k}}a^{\dagger}_{\boldsymbol{k},\lambda}a_{\boldsymbol{k},\lambda}$ and the Coulomb interaction between charges
\begin{equation}
H_{{\rm Charge}}=\sum_{l\neq m}\frac{1}{8\pi\varepsilon_{0}}\frac{q^2_{e}}{\left|\boldsymbol{x}^{(l)}-\boldsymbol{x}^{(m)}\right|}.
\end{equation}

The Heisenberg equation for the position of $l$th particle gives its velocity operator 
\begin{equation}
\boldsymbol{v}^{(l)}\equiv\dot{\boldsymbol{x}}^{(l)}=c\boldsymbol{\alpha}^{(l)}.
\end{equation}
The mechanical momentum of a charge particle in the Coulomb gauge is given by 
\begin{equation}
\boldsymbol{p}_{{\rm mech}}^{(l)}\equiv\boldsymbol{p}^{(l)}-q_{e}\boldsymbol{A}_{\perp}(\boldsymbol{x}^{(l)}).
\end{equation}
The Heisenberg equation for $\boldsymbol{p}_{{\rm mech}}^{(l)}$ gives the quantum Lorentz force equation
\begin{equation}
\frac{d}{dt}\boldsymbol{p}_{{\rm mech}}^{(l)}=\frac{i}{\hbar}\left[H,\boldsymbol{p}_{{\rm mech}}^{(l)}\right].
\end{equation}
The first term of $H$ gives 
\begin{align}
 & \frac{i}{\hbar}\left[c\boldsymbol{\alpha}^{(l)}\cdot\left(\boldsymbol{p}^{(l)}-q_{e}\boldsymbol{A}_{\perp}(\boldsymbol{x}^{(l)})\right),p_{{\rm mech,j}}^{(l)}\right]\nonumber \\
= & -\frac{i}{\hbar}q_{e}\left[c\boldsymbol{\alpha}^{(l)}\cdot\boldsymbol{p}^{(l)},A_{\perp,j}(\boldsymbol{x}^{(l)})\right]-\frac{i}{\hbar}q_{e}\left[c\boldsymbol{\alpha}^{(l)}\cdot\boldsymbol{A}_{\perp}(\boldsymbol{x}^{(l)}),p_{{\rm j}}^{(l)}\right]\nonumber\\
= & q_{e}\sum_{i}c\alpha_{i}\left[\partial_{j}A_{\perp,i}(\boldsymbol{x}^{(l)})-\partial_{i}A_{\perp,j}(\boldsymbol{x}^{(l)})\right]\nonumber\\
= & q_{e}\sum_{i}\sum_{k}v_{i}\epsilon_{jik}B_{\perp,k} = q_{e}\left[\boldsymbol{v}\times\boldsymbol{B}_{\perp}\right]_{j}.
\end{align}
Here, we see that no longitudinal magnetic field can enter the Lorentz force in the asymmetric representation. The second term of $H$ gives 
\begin{equation}
\frac{i}{\hbar}\left[H_{{\rm Charge}},p_{{\rm mech,j}}^{(l)}\right]=-q_{e}\partial_{j}V_{{\rm Charge}}=q_{e}E_{\parallel,j},
\end{equation}
where $V_{{\rm Charge}}$ is the Coulomb potential generated by other
charges. The third term gives (also see Chap. III B in~\cite{cohen1997photons})
\begin{equation}
\frac{i}{\hbar}\left[H_{M}^{T},p_{{\rm mech,j}}^{(l)}\right]=q_{e}E_{\perp,j},
\end{equation}
where we have used the plane-wave expansions
\begin{align}
\boldsymbol{A}_{\perp}(\boldsymbol{x}) &\! =\!\!\int\!\!\! d^{3}k\!\!\sum_{\lambda=1,2}\!\!\sqrt{\frac{\hbar}{2\varepsilon_{0}\omega_{\boldsymbol{k}}(2\pi)^{3}}}\!\left[a_{\boldsymbol{k},\lambda}\epsilon^{\mu}(\boldsymbol{k},\lambda)e^{i\boldsymbol{k}\cdot\boldsymbol{x}}\!+\!{\rm h.c.}\right]\!,\\
\boldsymbol{E}_{\perp}(\boldsymbol{x}) &\! =\!i\!\!\!\int\!\!\! d^{3}k\!\sqrt{\frac{\hbar\omega_{\boldsymbol{k}}}{2\varepsilon_{0}(2\pi)^{3}}}\left[a_{\boldsymbol{k},1}\boldsymbol{\epsilon}(\boldsymbol{k},1)\!+\!a_{\boldsymbol{k},2}\boldsymbol{\epsilon}(\boldsymbol{k},2)\right]e^{i\boldsymbol{k}\cdot\boldsymbol{x}}\!+\!{\rm h.c.}
\end{align}

We recovers the classical Lorentz force equation by regrouping all the preceding results
\begin{equation}
\frac{d}{dt}\boldsymbol{p}_{{\rm mech}}^{(l)}=q_{e}\left[\boldsymbol{E}(\boldsymbol{x}^{(l)})+\boldsymbol{v}\times\boldsymbol{B}_{\perp}(\boldsymbol{x}^{(l)})\right].
\end{equation}
Its counter part in the symmetric reorientation can be obtained by simply perform a dual rotation
\begin{align}
\frac{d}{dt}\boldsymbol{p}_{{\rm mech}}^{(l)}= & \tilde{q}_{e}\left[\tilde{\boldsymbol{E}}(\boldsymbol{x}^{(l)})+\boldsymbol{v}\times\tilde{\boldsymbol{B}}_{\perp}(\boldsymbol{x}^{(l)})\right]\nonumber\\
& +c\varepsilon_0\tilde{q}_{m}\left[c\tilde{\boldsymbol{B}}(\boldsymbol{x}^{(l)})-\frac{1}{c}\boldsymbol{v}\times\tilde{\boldsymbol{E}}_{\perp}(\boldsymbol{x}^{(l)})\right].
\end{align}

We note that the symmetrized representation is exactly equivalent to the original asymmetric one. Thus, the energy-momentum conservation conditions do not change. All conserved quantities in the symmetrized representation can be obtained from their asymmetric counterparts via the dual transformation, such as the canonical momentum $\tilde{\boldsymbol{p}}^{l}=\boldsymbol{p}^l_{\rm mech} -\tilde{q}_e\tilde{\boldsymbol{A}}(\boldsymbol{x}^{(l)})-\varepsilon_0\tilde{q}_m\tilde{\boldsymbol{C}}(\boldsymbol{x}^{(l)})$.

\section{QED with two independent gauge fields\label{sec:Hcharge2}}
In this section, we assume that the two gauge fields $\tilde{A}^{\mu}$ and $\tilde{C}^{\mu}$ are independent with each other and quantize them separately. The corresponding charges (sources) for these two gauge fields are $\tilde{q}_e$ and $\tilde{q}_m$, respectively~\cite{singleton1996electromagnetism}. Then, we check the photon-mediated charge-charge interaction. 

To quantize the two gauge fields, we assume the following equal-time
commutation relations 
\begin{align}
[\tilde{A}^{\mu}(\boldsymbol{x},t),\tilde{\pi}_{A}^{\nu}(\boldsymbol{x}',t)] & =i\hbar cg^{\mu\nu}\delta^{3}(\boldsymbol{x}-\boldsymbol{x}'),\\{}
[\tilde{C}^{\mu}(\boldsymbol{x},t),\tilde{\pi}_{C}^{\nu}(\boldsymbol{x}',t)] & =i\hbar cg^{\mu\nu}\delta^{3}(\boldsymbol{x}-\boldsymbol{x}'),\\{}
[\tilde{A}^{\mu}(\boldsymbol{x},t),\tilde{\pi}_{C}^{\nu}(\boldsymbol{x}',t)] & =[\tilde{C}^{\mu}(\boldsymbol{x},t),\tilde{\pi}_{A}^{\nu}(\boldsymbol{x}',t)]=0,\\{}
[\tilde{A}^{\mu}(\boldsymbol{x},t),\tilde{A}^{\nu}(\boldsymbol{x}',t)] & =[\tilde{\pi}_{A}^{\mu}(\boldsymbol{x},t),\tilde{\pi}_{A}^{\nu}(\boldsymbol{x}',t)]=0,\\{}
[\tilde{C}^{\mu}(\boldsymbol{x},t),\tilde{C}^{\nu}(\boldsymbol{x}',t)] & =[\tilde{\pi}_{C}^{\mu}(\boldsymbol{x},t),\tilde{\pi}_{C}^{\nu}(\boldsymbol{x}',t)]=0.
\end{align}
Their plane-wave expansions are given by
\begin{align}
\tilde{A}^{\mu} & \!=\!\!\int\!\! d^{3}k\sum_{\lambda=0}^{3}\!\sqrt{\frac{\hbar}{2\varepsilon_{0}\omega_{\boldsymbol{k}}(2\pi)^{3}}}\left[a_{\boldsymbol{k},\lambda}\epsilon^{\mu}(\boldsymbol{k},\lambda)e^{i\boldsymbol{k}\cdot\boldsymbol{x}}\!+\!{\rm h.c.}\right],\\
\tilde{C}^{\mu} & \!=\!\!\int\!\! d^{3}k\sum_{\lambda=0}^{3}\!\!\sqrt{\frac{c^{2}\hbar}{2\varepsilon_{0}\omega_{\boldsymbol{k}}(2\pi)^{3}}}\left[b_{\boldsymbol{k},\lambda}\epsilon^{\mu}(\boldsymbol{k},\lambda)e^{i\boldsymbol{k}\cdot\boldsymbol{x}}\!+\!{\rm h.c.}\right],\\
\tilde{\pi}_{A}^{\mu} & \!=\!i\!\!\int\!\! d^{3}k\sum_{\lambda=0}^{3}\!\!\sqrt{\frac{\hbar\omega_{\boldsymbol{k}}}{2\mu_{0}(2\pi)^{3}}}\left[a_{\boldsymbol{k},\lambda}\epsilon^{\mu}(\boldsymbol{k},\lambda)e^{i\boldsymbol{k}\cdot\boldsymbol{x}}-{\rm h.c.}\right],\\
\tilde{\pi}_{C}^{\mu} & =i\int d^{3}k\sum_{\lambda=0}^{3}\sqrt{\frac{\varepsilon_{0}\hbar\omega_{\boldsymbol{k}}}{2(2\pi)^{3}}}\left[b_{\boldsymbol{k},\lambda}\epsilon^{\mu}(\boldsymbol{k},\lambda)e^{i\boldsymbol{k}\cdot\boldsymbol{x}}-{\rm h.c.}\right],
\end{align}
where $a_{\boldsymbol{k},\lambda}$ and $b_{\boldsymbol{k},\lambda}$ have been interpreted as the annihilation operators of the electric and magnetic photons respectively~\cite{singleton1996electromagnetism} and they are independent with each other, i.e., $[a_{\boldsymbol{k},\lambda},b_{\boldsymbol{k},\lambda}]=[a_{\boldsymbol{k},\lambda},b^{\dagger}_{\boldsymbol{k},\lambda}]=0$.

Now, the Hamiltonian of the photons is given by $\tilde{H}_M= \tilde{H}_{M}^{T}+\tilde{H}_{M}^{L}+\tilde{H}_{M}^{S} =$ with
\begin{align}
\tilde{H}_{M}^{T}=&\!\! \int\!\! d^{3}k\hbar\omega_{\boldsymbol{k}}(a_{\boldsymbol{k},1}^{\dagger}a_{\boldsymbol{k},1}\!+\!a_{\boldsymbol{k},2}^{\dagger}a_{\boldsymbol{k},2}\!+\!b_{\boldsymbol{k},1}^{\dagger}b_{\boldsymbol{k},1}\!+\!b_{\boldsymbol{k},2}^{\dagger}b_{\boldsymbol{k},2}),\\
\tilde{H}_{M}^{L} =& \int d^{3}k\hbar\omega_{\boldsymbol{k}}(a_{\boldsymbol{k},3}^{\dagger}a_{\boldsymbol{k},3}+b_{\boldsymbol{k},3}^{\dagger}b_{\boldsymbol{k},3}),\\
\tilde{H}_{M}^{S}= & -\int d^{3}k\hbar\omega_{\boldsymbol{k}}(a_{\boldsymbol{k},0}^{\dagger}a_{\boldsymbol{k},0}+b_{\boldsymbol{k},0}^{\dagger}b_{\boldsymbol{k},0}).
\end{align}
Here, we see that the photonic density of states gets doubled. The interaction between the Maxwell and Dirac fields $\tilde{H}_{\rm int}=\tilde{H}_{{\rm int}}^{T}+\tilde{H}_{{\rm int}}^{L}+\tilde{H}_{{\rm int}}^{S}$ changes into
\begin{align}
\tilde{H}_{{\rm int}}^{T}+\tilde{H}_{{\rm int}}^{L} = & \!-\!\!\int\!\! d^{3}k\hbar\omega_{\boldsymbol{k}}\sum_{\lambda=1}^{3}\nonumber\\
&\left\{ \left[a_{\boldsymbol{k},\lambda}^{\dagger}\boldsymbol{\xi}_{e}(\boldsymbol{k})+b_{\boldsymbol{k},\lambda}^{\dagger}\boldsymbol{\xi}_{m}(\boldsymbol{k})\right]\cdot\boldsymbol{\epsilon}(\boldsymbol{k},\lambda)+{\rm h.c.}\right\},\\
\tilde{H}_{{\rm int}}^{S} & \!=\!\! \int\!\! d^{3}k\hbar\omega_{\boldsymbol{k}}\left\{ \left[a_{\boldsymbol{k},0}^{\dagger}\xi_{e,0}(\boldsymbol{k})\!+\!b_{\boldsymbol{k},0}^{\dagger}\xi_{m,0}(\boldsymbol{k})\right]\!+\!{\rm h.c.}\right\} .
\end{align}

We now require that both gauge field should satisfy the Lorenz condition. Then, the Gupta-Bleuler gauge condition changes into
\begin{align}
\left[a_{\boldsymbol{k},3}-a_{\boldsymbol{k},0}+\xi_{e,0}(\boldsymbol{k})\right]\left|\Phi\right\rangle & = 0,\\
\left[b_{\boldsymbol{k},3}-b_{\boldsymbol{k},0}+\xi_{m,0}(\boldsymbol{k})\right]\left|\Phi\right\rangle & = 0.
\end{align}

After eliminating the scalar photons via $\left\langle \Phi\right|\tilde{H}_{M}^{L}+\tilde{H}_{M}^{S}+\tilde{H}_{{\rm int}}^{S}\left|\Phi\right\rangle$, we now obtain the interaction between the charges,
\begin{equation}
  H_{\rm charge}= \!\int\!\! d^{3}x\!\!\int\!\! d^{3}x'\left[\frac{\tilde{\rho}_{e}(\boldsymbol{x})\tilde{\rho}_{e}(\boldsymbol{x}')}{8\pi\varepsilon_{0}|\boldsymbol{x}-\boldsymbol{x}'|}\!+\!\frac{(c\varepsilon_0)^2\tilde\rho_{m}(\boldsymbol{x})\tilde{\rho}_{m}(\boldsymbol{x}')}{8\pi\varepsilon_{0}|\boldsymbol{x}-\boldsymbol{x}'|}\right]. \label{eq:Coulomb-2}
\end{equation}
Here, we find that the interaction between $\tilde{q}_e$ charges is induced by $\tilde{A}_{\mu}$ gauge field (regular photons) and the interaction between $\tilde{q}_m$ charges is induced by $\tilde{C}_{\mu}$ gauge field (magnetic photons) and both interactions are of Coulomb form. However, the gauge fields cannot induce any interaction between two different charges. Thus, the introduction of the magnetic photons will not solve the $\tilde{q}_e-\tilde{q}_m$ interaction problem.

\section{QED with a generalized Lagrangian\label{sec:Lprime}}

In the interaction Lagrangian $\tilde{\mathcal{L}}_{\rm int}$ in Eq.~(\ref{eq:L_int}), the electric charge $\tilde{q}_e$ only interacts with the gauge field $\tilde{A}^{\mu}$ and the dual charge $\tilde{q}_m$ only interacts with the gauge field $\tilde{C}^{\mu}$. If we treat the two gauge fields as two independent fields, there will be no interaction between these two types of charges as expected. In principle, we can manually construct the following interaction Lagrangian density
\begin{equation}
\tilde{\mathcal{L}}^{\prime}_{{\rm int}}=-c\bar{\psi}\gamma_{\mu}\left[\tilde{q}_{e}(\tilde{A}^{\mu}-\tilde{C}^{\mu}/c)+\varepsilon_{0}q_{m}(\tilde{C}^{\mu}+c\tilde{A}^{\mu})\right]\psi,
\end{equation}
which is also invariant under a dual transformation. Then, interaction between $\tilde{q}_e$ and $\tilde{q}_m$ could be mediated by the two independent gauge fields. However, this interaction Lagrangian has its own problems. 
 
The Lagrangian density $\tilde{\mathcal{L}}^{\prime}_{{\rm int}}$ is not a scalar anymore, because $\tilde{q}_m$ is a pseudo-scalar and $\tilde{C}^{\mu}$ is a pseudo-vector. Via the Euler-Lagrange equation, we obtain the equations for the electric and magnetic fields
\begin{align}
\boldsymbol{\nabla}\cdot\tilde{\boldsymbol{E}} &   =\tilde{\rho}_{e}/\varepsilon_{0}+c\tilde{\rho}_{m},\\ \boldsymbol{\nabla}\cdot\tilde{\boldsymbol{B}} & =\tilde{\rho}_{m}-\tilde{\rho}_{e}/c\varepsilon_{0},\\
\boldsymbol{\nabla}\times\tilde{\boldsymbol{E}} & = -\frac{\partial}{\partial t}\tilde{\boldsymbol{B}}-\tilde{\boldsymbol{j}}_{m}+\tilde{\boldsymbol{j}}_{e}/c\varepsilon_{0}, \\
\boldsymbol{\nabla}\times\tilde{\boldsymbol{B}} & = \frac{1}{c^{2}}\frac{\partial}{\partial t}\tilde{\boldsymbol{E}}+\mu_{0}\tilde{\boldsymbol{j}}_{e}+\tilde{\boldsymbol{j}}_{m}/c,
\end{align}
which are not the Maxwell equations anymore. 

After eliminating the scalar photons, we obtain the following interaction Hamiltonian
\begin{equation}
\tilde{H}^{\prime}_{\rm charge} \!=\! \left\langle \Phi\right|\!\!\int\!\! d^{3}x\!\!\int\!\! d^{3}x'\frac{\tilde{\rho}_{e}(\boldsymbol{x})\tilde{\rho}_{e}(\boldsymbol{x}')+c^{2}\varepsilon_{0}^{2}\tilde{\rho}_{m}^{\dagger}(\boldsymbol{x})\tilde{\rho}_{m}(\boldsymbol{x}')}{4\pi\varepsilon_{0}|\boldsymbol{x}-\boldsymbol{x}'|}\left|\Phi\right\rangle. \label{eq:Coulomb-3}
\end{equation}
Here, we see that the obtained charge-charge interaction is still of Coulomb-type and no interaction between electric charge and the dual charge has been obtained. Compared to Eq.~(\ref{eq:Coulomb-2}), the Coulomb interaction is doubled. 

\bibliography{main}
\end{document}